# Advantages of high-speed technique for quantum key distribution; reply to quant-ph/040750

**J. C. Bienfang, Charles W. Clark, Carl J. Williams**

*National Institute of Standards and Technology, 100 Bureau Dr., Gaithersburg, MD 20899*
*bienfang@nist.gov*

**E. W. Hagley, Jesse Wen**

*Acadia Optronics LLC, 13401 Valley Drive, Rockville, MD 20850*

**Abstract:** We respond to a comment on our high-speed technique for the implementation of free-space quantum key distribution (QKD). The model used in the comment assigns inappropriately high link losses to the technique in question; we show that the use of reasonable loss parameters in the model invalidates the comment's main conclusion and highlights the benefits of increased transmission rates.

A recent comment in this forum by Hughes and Nordholt [1] (hereafter HN) discusses a paper published by Bienfang *et al.* [2] (hereafter B) on quantum key distribution (QKD). HN suggests that substantial conclusions of B are misleading, and that its implications for QKD are obscure. These assertions disregard the relevance and scope of the work presented in B.

The generation of sifted key is the primary function of the physical layer of any single-photon QKD system, and B presents a scalable technique that significantly increases the capacity of the physical layer. This technique is based on the adaptation of clock synchronization technology from modern telecommunications for use in QKD, and dramatically increases the transmission rate of the quantum channel. As an experimental demonstration of this technique, B reports the continuous generation of sifted quantum cryptographic key over a 730 m free-space link at rates of 1 Megabit/second, with an error rate of 1.1%. The sifted rate performance reported in B constitutes an increase in throughput over previously published reports by two orders of magnitude, and we stand by this conclusion. As stated in HN, error correction and privacy amplification impose an additional overhead in the final throughput. However, the relative gain to be had from increased transmission rate is unaffected.

HN presents an extrapolation of the results of B to full QKD over a 10 km link, and asserts that the higher transmission rate would result in no improvement beyond that which has been demonstrated, for example, in ref. [3]. To arrive at this conclusion, HN imputes to B a link loss factor of –29 dB that is derived from treating the optical propagation of B in the asymptotic limit, where the intensity scales directly as the inverse-square of the distance. However, previous results [3] demonstrate that it is possible to design an optical system that experiences losses in the range of –13 dB to –19 dB over a 10 km link. If we assume these same loss factors, we find that the analysis presented in HN would predict two orders of magnitude more secret key than reported in ref. [3]. This result is in agreement with straightforward scaling arguments, and disaffirms the statement in HN that the method of B would offer no advantage for QKD. In fact, we expect that optimization for higher transmission rates will continue to produce significant increases in throughput.

The argument of HN concerning the relationship between rate and reach of QKD deserves careful consideration, as it states that there is no advantage in reach to be gained by increasing the transmission rate. In terms of telecommunications engineering considerations, any usable QKD system will be specified by the link length, $L$, and a requirement for the rate, $R$, of generation of secret key. The main barrier in extending the length to $L' > L$, while keeping $R$ constant, is the greater link loss. As HN notes, there is a set of link parameters beyond which no secret key can be generated. Provided that this limit is not exceeded, any additional losses incurred by extending the link length that reduce $R$ can be compensated for by increasing the transmission rate. HN's conclusion is based on taking $L$ to be the length at which no usable secret key is obtained ($R = 0$). At that point no change in transmission rate will affect the production of key, because there is no key produced. This is correct as a logician's exercise in *reductio ad absurdum*, but is not useful as a design criterion.

______________________________________________________________________________

## References and links

______________________________________________________________________________